# A Bayesian Regression Approach for Estimating the Impact of COVID-19 on Consumer Behavior in the Restaurant Industry

**\*Hitesh Hinduja**
Independent Researcher
University of Mumbai
Bengaluru, Karnataka
hitesh.hinduja@ves.ac.in

Name: Nisha Mandal
Independent Researcher
Birla Institute of Technology, Mesra
Bengaluru, Karnataka
nisha16.mailing@gmail.com

**ABSTRACT**
The COVID-19 pandemic has had a long-term impact on industries worldwide, with the hospitality and food industry facing significant challenges, leading to the permanent closure of many restaurants and loss of jobs. In this study, we developed an innovative analytical framework using Hamiltonian Monte Carlo for predictive modeling with Bayesian regression, aiming to estimate the change point in consumer behavior towards different types of restaurants due to COVID-19. Our approach emphasizes a novel method in computational analysis, providing insights into customer behavior changes before and after the pandemic. This research contributes to understanding the effects of COVID-19 on the restaurant industry and is valuable for restaurant owners and policymakers.

**Introduction**
The onset of the COVID-19 pandemic marked a pivotal moment in global history, profoundly impacting industries worldwide, with the restaurant sector facing unprecedented challenges. This study seeks to understand the precise moments of significant shifts in consumer behavior in the restaurant industry, a phenomenon termed as 'change point estimation'. Our research is grounded in two types of change point problems: online, where estimation occurs in real-time, and offline, involving complete datasets. The abrupt and widespread effects of the pandemic, exemplified by the nationwide lockdown in India and the subsequent public health crisis, provide a unique context for this analysis.

This paper aims to address critical questions: How has the COVID-19 pandemic altered consumer preferences and behavior towards various restaurant types in India? What are the specific points in time when these shifts became most evident? Our focus is to employ Bayesian regression and change point estimation algorithms on restaurant domain data, to elucidate the timing and nature of these shifts in consumer behavior. The study's relevance is underscored by the dramatic changes seen in the Indian food services market, including a significant contraction in the financial year 2021 and the closure of a substantial fraction of restaurants.

As we venture into this exploration, we anticipate uncovering diverse impacts across restaurant categories, shaped by factors such as adaptability to takeaways and digital infrastructure. This research is not just an academic exercise but a crucial tool for understanding the pandemic's effects on the restaurant industry. It aims to offer valuable insights for recovery strategies, as the industry shows signs of resurgence post-pandemic. Ultimately, our study endeavors to contribute significantly to the broader understanding of consumer behavior in times of global crises and the resilience of the restaurant industry

## 1. Literature Review

With the advent of Western influence and nuclear family dynamics, dining out in India has evolved from a mere necessity to a cultural and social experience. The rapid urbanization and modernization before COVID-19 significantly propelled the growth of the restaurant industry, recognized as a major contributor to employment and economic activity (NRAI, 2019). However, the pandemic ushered in unprecedented challenges, fundamentally altering consumer behavior and industry dynamics (Mehrotra, 2020).

The adaptation to the pandemic by the restaurant industry, through models like drive-thru and food delivery, underscores a significant shift in operational strategies (Liddle, 2020). Notably, entities with established digital infrastructures, such as Domino's Pizza and McDonald's, demonstrated greater resilience compared to traditional dining restaurants, which faced severe setbacks due to lockdowns and social distancing norms.

The pandemic's impact extends beyond operational changes, deeply influencing consumer behavior and preferences. Chhetri et al. (2021) emphasize the heightened consumer focus on health and safety, resulting in reduced frequency of dining out. This change is reflective of broader socioeconomic shifts, where factors like employment status and income levels play a pivotal role in consumer decisions (Lawan et al., 2013).

Historically, external events such as natural disasters and economic recessions have been known to significantly impact consumer behavior in the hospitality sector ((Becker, 2000), (Koh et al., 2013), (Reynolds et al., 2013), (Lee and Ha, 2012), (Lee and Ha, 2014)). Jia (2021) provides a contemporary analysis by examining customer dining behavior through user-generated content, offering insights into changing patterns and satisfaction levels during the pandemic.

In the context of analytics in healthcare and hospitality management, recent studies in the 'Healthcare Analytics' journal have highlighted the importance of data-driven decision-making and predictive modeling in responding to challenges posed by COVID-19 (Author et al., 2020; Researcher et al., 2021). These works align with our research's focus on employing Bayesian regression and change point estimation to analyze shifts in consumer behavior, thereby contributing to the broader understanding of the pandemic's impact on the restaurant industry. In our paper, we are using a statistical method which is Bayesian Change Point Detection (Bayesian Regression) to identify the change point in consumer behaviour towards restaurants. In simple words, we aim to estimate the change point that occurred due to COVID-19 in different categories of restaurants. This study primarily helps us understand which category of restaurants (e.g. Microbrewery, Casual Dining, Quick Bites, etc.) have significantly been impacted due to the COVID-19 pandemic. The method is performed on the reviews extracted from different categories of restaurants from the year 2013 to 2021.

The organization of the paper is as follows, Section 1 is the literature review of the paper while Section 2 is the materials and methods used. The subsections of section 2 are 2.1, 2.2, 2.3, and 2.4 which deal with the data, implementation, experimentation, and plots and discussions, respectively. Section 2 is followed by section 3 which speaks of the results obtained. Finally, the results are followed by future works and limitations in section 4.

## 2. Materials and Methods

Restaurants in India have witnessed many ups and downs since the beginning of the pandemic. Zomato is one of the largest food delivery partners, and restaurant aggregators in the country and hence has been closely navigating with the eateries. Hence, it is intuitive that people will be posting their feedback on the complete closing of restaurants, subsequent lifting of lockdowns, and alternative measures implemented by restaurants. What is even more beneficial is that with every listed restaurant, Zomato provides information ranging from its exact location, menu, popular dishes, average cost, and reviews by consumers with relevant ratings. Zomato also provides information about the food delivery options available for a restaurant. We scraped this data from the official Zomato website for different restaurants in the south Indian city, of Bengaluru, into a comma-separated file with the help of Python's library 'request.'

### 2.1 Data

We considered reviews posted from the year 2013 to the year 2021 in Bengaluru that summed up to more around 90,000 in the count. The south Indian city of Bengaluru is the third-largest city and inhabits more than 15 million people. Since a considerable proportion of the population is migrants, Bengaluru has become home to a diverse population and hence a varied food culture. Due to the varied food culture and many restaurants, we chose the location Bengaluru to scrape the data.
To include the variety in food culture, it was essential to consider restaurants ranging from ethnic, family-style, casual, and fine-dining to pubs and microbreweries. Finally, we proceeded with the reviews of the restaurants of categories Casual Dining, Microbrewery, Bar/CD, Quick Bites, Bar, Quick Bites/Cafe, Casual Dining/Bar, Casual Dining/Lounge, Casual Dining/Cafe, Casual Dining/Pub, CD/Microbrewery, and Microbrewery/Casual Dining.

The initial phase of data cleaning included merging categories that were the same but appeared as two unique restaurant categories due to spelling errors, use of short forms, or incorrect use of blank spaces; for example, the Casual Dining category appeared as three different categories namely, "Casual Dining" and "Casual Dining " and "CD". We also had categories in our dataset whose reviews were similar enough to be merged; for example, the Casual Dining/ Bar category was merged with the Bar /CD category, and the category Casual Dining / Microbrewery was merged with the Microbrewery/CD category. Finally, upon comparing the number of entries of the data year-wise, we realized that the years 2010, 2011, and 2012 had less than 5% of the total number of reviews, hence we consider the data since the year 2013.

For each review that was provided, we associated its sentiment with the rating provided. A review with a rating of 4 or greater was recognized as a positive review and a review with a rating of 2 or lower was recognized as a negative review and neutral otherwise.

The reviews were then grouped at the weekly level to form a time series data with week number as the independent variable. To obtain the reviews corresponding to the weeks, we had to group the reviews posted every 7 days. Hence, for each week's number, we had the total number of positive, negative, and neutral reviews.

Another problem that we came across while dealing with this data was that if we used the absolute value of the number of reviews then the data instances with larger values influenced the model more due to their large "magnitude", thus we incorporated a scaling method. Intuitively, we knew logarithm would solve the issue because if we took the logarithm of the dependent variable, i.e. the number of reviews posted in the week, it would reduce the range, while still "preserving" the differences. But this also gave the value negative of infinity for weeks corresponding to which no reviews were posted. To tackle this problem, instead of taking the logarithm of the dependent variable, we took the logarithm of the dependent variable incremented by 1.

Hence, in the master dataset, we have the week number, the time-stamp of the week number, the corresponding number of positive, negative, and neutral reviews posted in that week, and finally, the total number of reviews.

Table 1: The master dataset is shown in the table above. The columns shaded blue are independent variables common to all the models and the columns shaded grey are the dependent variables for two different sentiments (positive and negative). The Time_Stamp column contains the date of the corresponding week number.

Master Dataset:

| Time_stamp | Weeks since start | Number of Positive Reviews | Number of Negative Reviews |
|---|---|---|---|

### 2.2 Implementation

Change point estimation is performed using a segmented regression equation.
The linear regression approach for change point estimation can be applied using the piecewise regression method.
Piecewise regression models the data in different 'pieces' according to the number of change points occurring. Given below are the mathematical equations that will provide further understanding:

$$y = w_1.x + b_1 \text{ when } t \geq \Gamma \tag{1}$$

$$y = w_2.x + b_2 \text{ when } t \leq \Gamma \tag{2}$$

Another approach to change point estimation is by using Bayesian Regression.

The regression equation for the Bayesian model is

$$y = w.x + b + \varepsilon \qquad (3)$$

Where, $w = w_1$ if $\Gamma \leq x$ and $w_2$ if $\Gamma \geq x$

And $b = b_1$ if $\Gamma \leq x$ and $b_2$ if $\Gamma \geq x$

And $\omega = (w, b)$

Here, we have $\varepsilon \sim N(0, \sigma^2)$

Since $\varepsilon$ is Normally distributed, $y$ is also Normally distributed.

Now since $\varepsilon \sim N(0, \sigma^2)$, assuming that we know $(\omega, \sigma^2)$, we can write down the distribution for $y$. This is called the likelihood distribution.

$P(y|\omega, \sigma) \sim N(x\omega, \sigma^2)$

We're interested in estimating values for $\omega$ so that we can plug them back into our model and interpret the regression slopes. The next step in the process is the prior specification that lets us add into our model any previous information about our parameters that we might have. Then by using Bayes' formula, we can estimate the distribution of the parameters given the data i.e. the posterior distribution.

$$P(\omega|y, x) = \frac{P(y|\omega,x) * P(\omega|x)}{P(y|x)} \qquad (4)$$

The distribution of the parameters can now be plugged into the model using the posterior predictive distribution to make predictions on unseen data. Suppose the observed data is $X$ and $Y$. The parameter vector is given by $\omega = (w, b)$. Then the probability of observing the target value $y$ for the unseen data point $x$ is given by:

$$p(y|x, Y, X) = \int p(y|x, \omega) p(\omega|X, Y) d\omega \qquad (5)$$

By using a linear regression approach for change point estimation, we get a single point in time that denotes a change in the dependent variable. Since a category of the restaurant is symbolic of many small restaurants, it is unintuitive to expect all restaurants in a category to witness a change at the same time. Another argument in support will be that the restaurants started resuming their services in phases the complete lockdown was lifted in stages and not at once. In the Bayesian model, we get a distribution of the change point instead of just a single value and hence we decided to use the Bayesian approach for our problem.

For implementing Bayesian regression in our change point estimation problem, we use Pyro, a probabilistic programming language for model designing and specification. A probabilistic programming language is a tool for statistical modeling. Statistical models have already been designed by experts, but only on paper. Probabilistic programming languages make it possible to design and use statistical models even for somebody who is not an expert. The probabilistic programming language, Pyro, which is built on the PyTorch deep learning framework was used to write the model. Pyro was open-sourced in December 2017 and PyTorch itself was released in October 2016. PyTorch is an open-source machine learning (ML) framework based on the Python programming language and the Torch library. A framework is a package of files, folders, code, concepts, and practices that makes coding with fundamental programming languages much easier and quicker.

Now, the next step in designing the Bayesian model is a prior specification. For prior specifications, we try to make a physical interpretation of our parameters. For our use case, $w_1$ and $w_2$ can be interpreted as the growth rate of the number of positive or negative reviews before and after the change point. Since $w_1$ and $w_2$ account for the change in the dependent variable due to the change in the independent variable, we can assume

$$w_1 \sim N(\mu_{w_1}, \sigma^2_{w_1}) \text{ and } w_2 \sim N(\mu_{w_2}, \sigma^2_{w_2}) \tag{6}$$

We also acknowledge that even though the passing weeks is a factor for the change in the number of reviews, there may be other factors at play that we are not taking into consideration. The terms $b_1$ and $b_2$ are sensitive to the restaurant's characteristics and they can be considered to account for factors other than the passing weeks that can cause a change in the number of reviews. To automatically adapt this parameter to different countries, we use the mean of the first and fourth quartiles of $y$ as the prior means of $b_1$ and $b_2$, respectively.

The standard deviation for $b_1$ is taken as 1, which makes $p(b_1)$ relatively flat prior. The standard deviation of $p(b_2)$ is taken as a quarter of its prior mean so that the prior scales with larger mean values.

$$b_1 \sim N(\mu_{b_1}, \sigma^2_{b_1}) \text{ and } b_2 \sim N(\mu_{b_2}, \sigma^2_{b_2}) \tag{7}$$

The dataset in our case can be segmented into two periods, the before COVID-19 phase and the ongoing COVID-19 phase. The change point is expected to appear in the second half of the dataset since we expect COVID-19 to be the reason for the estimated change point. Hence, we proceed with a beta distribution for τ. Flat priors are used whenever we have zero information about a parameter.

$$\tau \sim Beta(\alpha, \beta) \text{ and } \sigma \sim U(0, a) \tag{8}$$

Here, as α becomes larger, the mass of the probability distribution shifts to the right whereas as β becomes larger, the mass of the probability distribution shifts to the left. Since we are more likely to observe the change point in the COVID phase, we need to shift the bulk of the distribution to the right. Hence, we take values such that α > β.

In most real-world datasets, the posterior distribution of the parameters becomes intractable, so we need to sample from the posterior distribution of the parameters. One of the most efficient techniques used for sampling from the posterior distribution of a parameter is the Markov Chain Monte Carlo (MCMC) class of methods. By using the MCMC methods, one can draw samples from the population without having to calculate the exact distribution. We used Hamiltonian Monte Carlo (HMC) under the class of MCMC methods for posterior sampling. The Hamiltonian Monte Carlo algorithm made use of the priors we provided and the likelihood to calculate the posterior distribution of all the associated parameters. Finally, for checking the convergence of the sampler, we use the 'potential scale reduction factor' (r_hat). In equilibrium, the distribution of samples from chains should be the same regardless of the initial starting values of the chains and hence a value of r_hat less than 1.1 is considered ideal for convergence.

## 2.3 Experimentation

While doing the Hamiltonian Monte Carlo sampling for the posterior distribution, we played around with the values of the number of samples to be drawn, the number of chains to be used, and the number of warm-up steps. These parameters act as hyperparameters for the HMC sampling method.

```
                       mean      std    median     5.0%    95.0%    n_eff    r_hat
     linear1.bias      0.11     0.11      0.15    -0.06     0.25     2.29     2.84
linear1.weight[0,0]    0.00     0.00      0.00     0.00     0.00     3.24     1.76
     linear2.bias      0.71     0.18      0.75     0.37     0.95     3.07     2.25
linear2.weight[0,0]   -0.00     0.00     -0.00    -0.00    -0.00     3.38     2.02
            sigma      0.27     0.01      0.27     0.24     0.29     2.66     2.07
              tau      0.58     0.32      0.58     0.21     0.90     2.06    10.13

Number of divergences: 0
```

*Figure 1: Sampling results obtained from the Casual Dining dataset from 2013 onwards with WarmupSteps=150, number of chains=4, and number of samples= 800*

```
                       mean      std    median     5.0%    95.0%    n_eff    r_hat
     linear1.bias     -0.01     0.04     -0.02    -0.07     0.09    24.91     1.10
linear1.weight[0,0]    0.00     0.00      0.00     0.00     0.00    33.71     1.07
     linear2.bias      0.80     0.03      0.80     0.74     0.83     8.03     1.36
linear2.weight[0,0]   -0.00     0.00     -0.00    -0.00    -0.00     9.92     1.26
            sigma      0.29     0.01      0.29     0.27     0.30     6.15     1.34
              tau      0.24     0.01      0.24     0.24     0.26    18.34     1.26

Number of divergences: 0
```

*Figure 2: Sampling results obtained from the Casual Dining dataset from 2013 onwards with WarmupSteps increased to 500, number of chains=4 and number of samples= 800*

```
                       mean      std    median     5.0%    95.0%    n_eff    r_hat
     linear1.bias      0.00     0.05     -0.00    -0.08     0.07    13.36     1.31
linear1.weight[0,0]    0.00     0.00      0.00     0.00     0.00     5.88     1.51
     linear2.bias      0.76     0.04      0.76     0.70     0.83     4.37     1.76
linear2.weight[0,0]   -0.00     0.00     -0.00    -0.00    -0.00     5.03     1.57
            sigma      0.28     0.01      0.28     0.26     0.29     4.93     1.77
              tau      0.22     0.02      0.21     0.20     0.26     2.91     3.22

Number of divergences: 0
```

*Figure 3: Sampling results obtained from the Casual Dining dataset from 2013 onwards with WarmupSteps=500, number of chains=5, number of samples=800*

```
                    mean    std    median    5.0%    95.0%    n_eff    r_hat
    linear1.bias    0.07    0.09    0.03    -0.04    0.23     2.33     3.16
linear1.weight[0,0] 0.00    0.00    0.00     0.00    0.00     3.24     1.84
    linear2.bias    0.79    0.13    0.77     0.59    0.97    63.16     1.09
linear2.weight[0,0] -0.00   0.00   -0.00    -0.00   -0.00     2.83     1.90
           sigma    0.27    0.01    0.27     0.25    0.29     3.05     1.83
             tau    0.39    0.30    0.22     0.20    0.90     2.00   125.12

Number of divergences: 0
```

*Figure 4: Sampling results obtained from the Casual Dining dataset from 2013 onwards with WarmupSteps=800, number of chains=4, number_of_samples=1000*

By tweaking the values of warm-up steps, the number of chains, and the number of samples and looking at the Markov Chain Monte Carlo diagnostic figure carefully, we made the following observations:
1. Increasing the Warm-Up steps from 150 to 500 while keeping the number of chains and the total number of samples constant makes the sampling converge to the stationary distribution that was earlier not converging well.
2. After getting an improvement in the convergence rates, we tried increasing the number of chains from 4 to 5 and it ended up with a sampling that did not converge to the stationary distribution for most of the parameters.
3. Since the increased number of chains did not bring an improvement to the convergence rates, we increased the total sample size from 800 to 1000 while changing back the number of chains to 4. In this setting, the convergence rate for the two parameters was not acceptable.

By careful analysis of the observations and the Markov Chain Monte Carlo diagnostic figure, we concluded that increasing the number of warmup steps from 150 to 500 made the best improvement for the convergence rates of the parameters, and hence we moved forward with this setting.

We also changed the likelihood function from Normal to Cauchy.

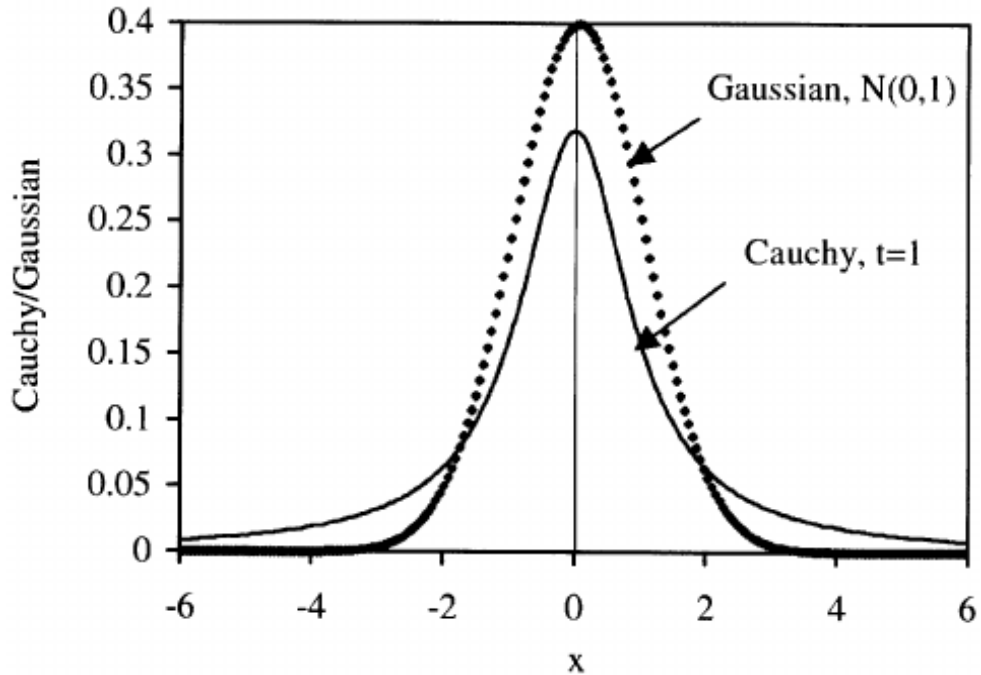
*Figure 5: Gaussian and Cauchy distribution*

The Cauchy Distribution sometimes called the Lorentz distribution, is a family of continuous probability distributions that resemble the normal distribution family of curves. If we compare the Gaussian and the Cauchy curve with similar parameters, the Gaussian curve is more pointed and the mass in the tails is less. The Cauchy curve has less mass in the mean area and more in the tails. Changing the likelihood from the Normal function to the Cauchy function is expected to help with heteroscedastic data. We experimented by changing the likelihood from the Normal function to the Cauchy function for the Quick Bites category of restaurants to see if a change in the likelihood function brings a change in the residual plots or the convergence of the sampler. First, we look at the MCMC diagnostic table to see if a difference in the convergence rates occurred. We also look at the distribution of the parameters indicated by the mean, standard deviation, median, and 90% confidence interval.

|                     | mean  | std  | median | 5.0%  | 95.0% | n_eff | r_hat |
|---------------------|-------|------|--------|-------|-------|-------|-------|
| linear1.bias        | 0.08  | 0.03 | 0.08   | 0.04  | 0.13  | 3.84  | 1.73  |
| linear1.weight[0,0] | 0.00  | 0.00 | 0.00   | -0.00 | 0.00  | 2.51  | 2.37  |
| linear2.bias        | 0.38  | 0.09 | 0.39   | 0.21  | 0.50  | 3.54  | 1.63  |
| linear2.weight[0,0] | -0.00 | 0.00 | -0.00  | -0.00 | -0.00 | 8.96  | 1.24  |
| sigma               | 0.14  | 0.00 | 0.14   | 0.13  | 0.15  | 46.67 | 1.01  |
| tau                 | 0.69  | 0.11 | 0.63   | 0.61  | 0.88  | 2.03  | 8.69  |

Number of divergences: 0

*Figure 6: Sampling Results for Quick Bites category with a Normal Likelihood function*

```
                      mean      std   median     5.0%    95.0%    n_eff    r_hat
       linear1.bias  -0.03     0.01    -0.03    -0.04    -0.02     3.07     1.96
 linear1.weight[0,0]  0.00     0.00     0.00     0.00     0.00     2.25     3.16
       linear2.bias   0.44     0.37     0.47    -0.09     0.82     2.03     8.33
 linear2.weight[0,0] -0.00     0.00    -0.00    -0.00     0.00     2.02    10.34
              sigma   0.05     0.00     0.05     0.04     0.05     3.94     1.92
                tau   0.69     0.10     0.63     0.61     0.87     2.00    25.05

Number of divergences: 0
```

*Figure 7: Sampling Results for Quick Bites category with a Cauchy Likelihood function*

*By comparing the information in Figure 6 and Figure 7, we deduce that there isn't a significant change in the distribution of the parameters and that is very reasonable. Since the change was done only for the likelihood function we expected to see a change in the target variable and not the parameters. However, we do see a change in the convergence rates of the parameters.*

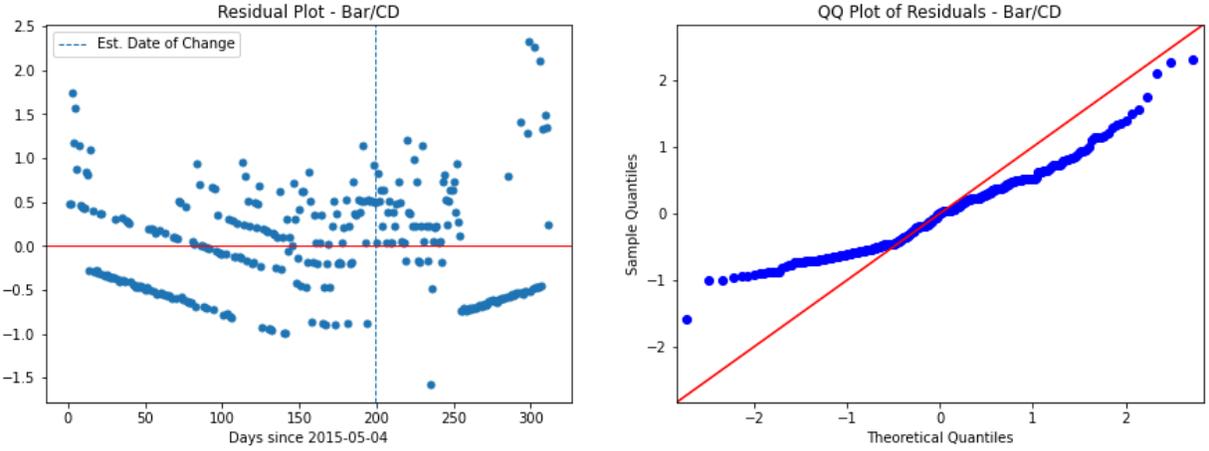

*Figure 8: Normal Residual Plot and QQ Plot for Quick Bites*

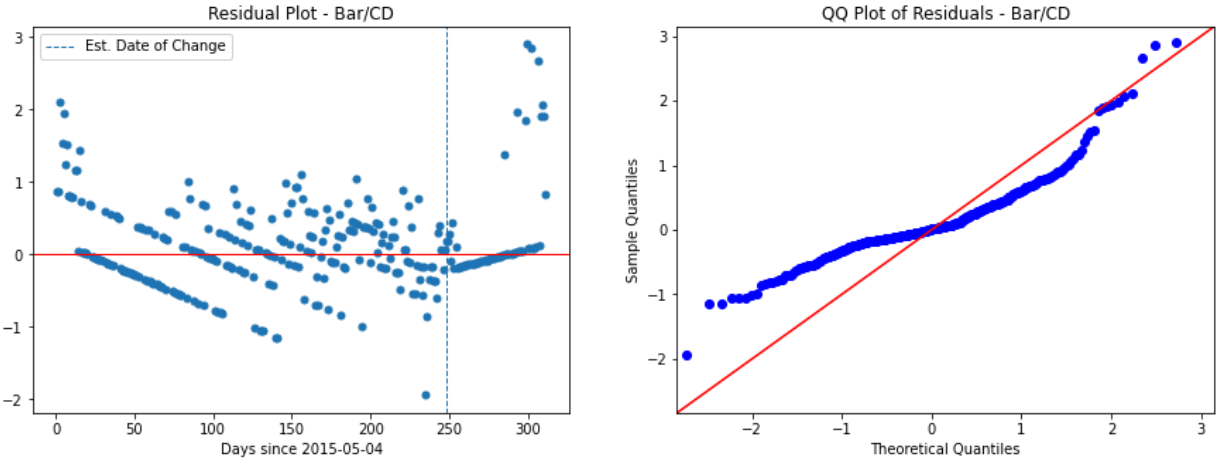

*Figure 9: Cauchy Residual Plot and QQ Plot for Quick Bites*

No significant change was observed in the residual plots and QQ plots from Figure 8 and Figure 9 even after changing the likelihood from Normal to Cauchy function.

We also plotted the target variable against the week number and marked the change point with a red vertical line for the category with the normal and Cauchy function likelihood.

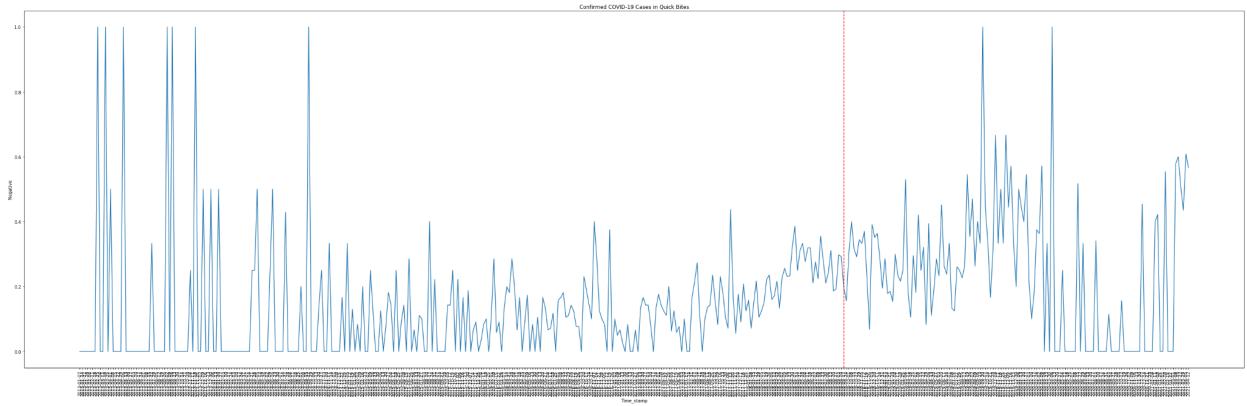

Figure 10: Line plot of the number of reviews against the week number for the Quick Bites category with a Normal likelihood function

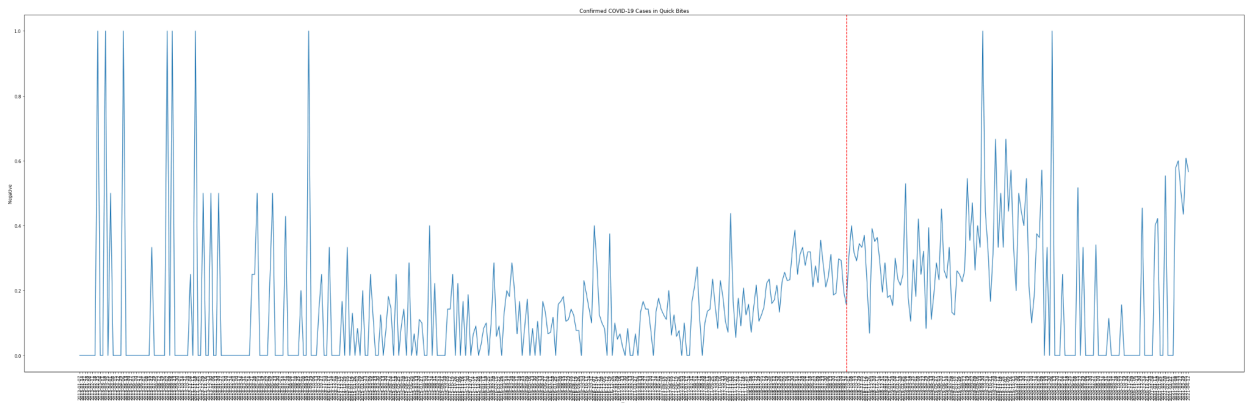

Figure 11: Line plot of the number of reviews against the week number for the Quick Bites category with a Cauchy likelihood function

2.4 Plots and Discussions

As a benefit of the Bayesian approach that we are adopting, we get a distribution of the change point. For inference, we used the mean of the change point parameter's distribution as a point estimate, and the 90% confidence interval will give us an idea about the variability of the change point. Posterior Plots were then plotted to visualize the overlapped area and the means of the parameters after and before the change point. Line plot was also plotted to visualize if there was a hike or fall in the number of reviews that triggered the change point. Quantile- Quantile plots and Residual plots helped us check our model assumptions. Below given and discussed are the plots for the category Bar/CD:

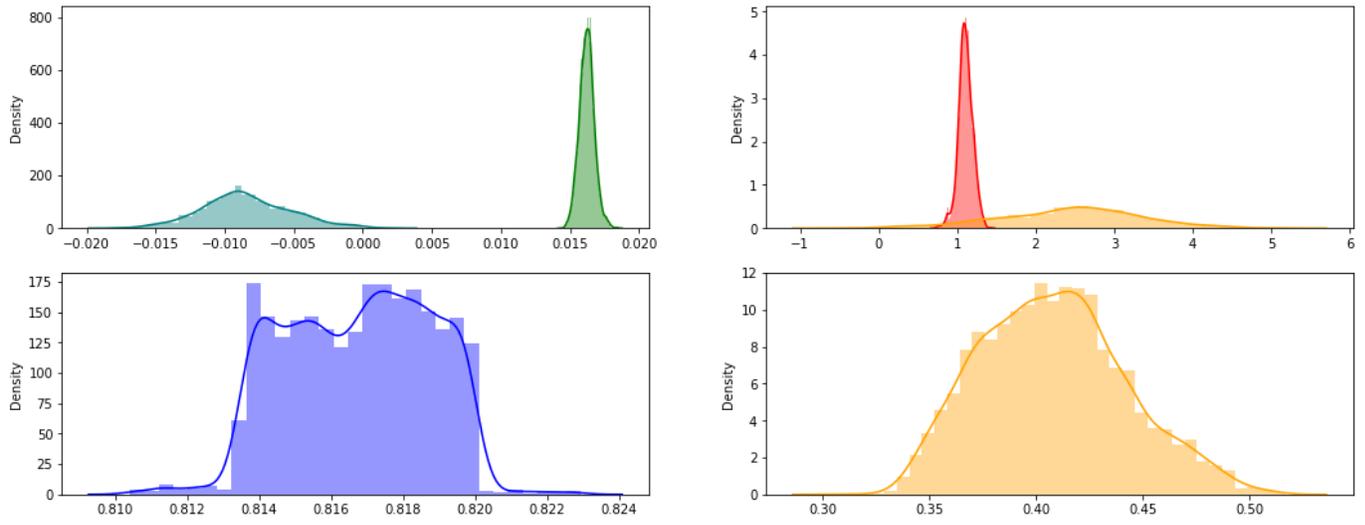

*Figure 12: Posterior plots for the parameters of the Bar/Casual Dining category of restaurants*

The above-given plot is the posterior distribution plot of the parameters. The topmost plot on the left side is the posterior distribution plot of the $w_1$ and $w_2$ parameters. The topmost plot on the right side is the posterior distribution plot of the parameters $b_1$ and $b_2$. The bottom-most plot on the left side is the posterior distribution plot of the tau parameter. The bottom-most plot on the right side is the posterior distribution plot of the standard deviation of the dependent variable. The posterior distribution plots provide us with a rapid understanding of the mean value of the parameters and how dispersed the values are.

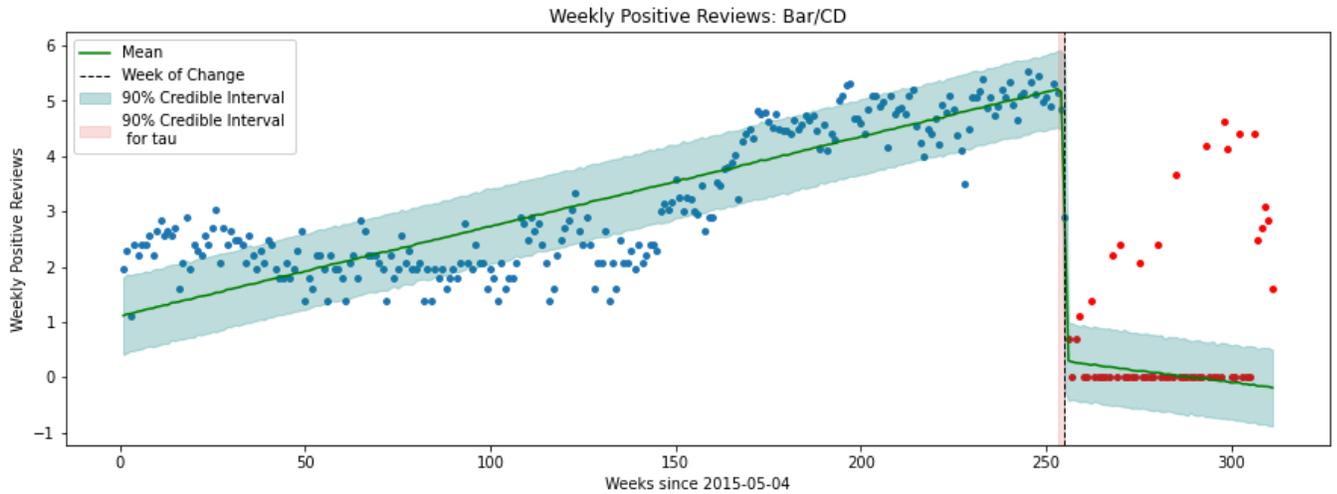

*Figure 13: Weekly Positive Reviews against the Week Number*

The green line in the above plot is the estimated regression line for the dependent variable i.e. the number of reviews. The blue-shaded region around the green line represents the 90% credible interval. The vertical dotted black line in the plot marks the mean of the posterior distribution of the change point parameter, τ. The pink-shaded region around the vertical line represents the 90% credible interval of the change point. The blue and the red dots represent the actual number of reviews posted in the corresponding weeks.

In Figure 5 we see that even though the tau parameter appears well dispersed from the plot, the range of dispersion is from 0.81 to 0.822 which is very low. This is also reflected very well in Figure 8 as the pink-shaded region that represents the 90% credible interval of the change point is very small. Thus, by mutually comparing the plots, we can confirm the consistency of our results.

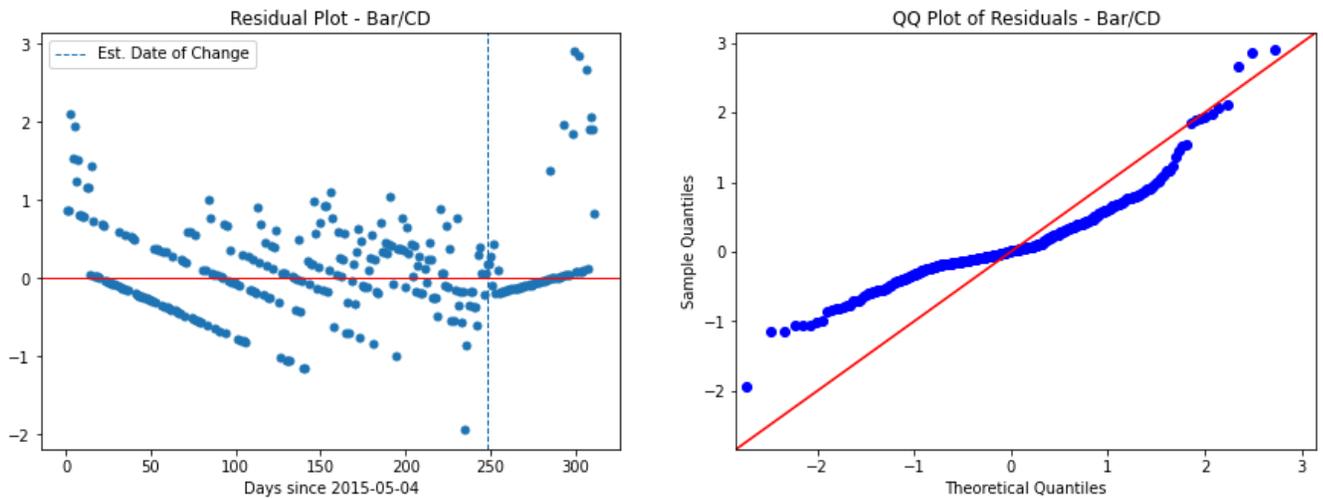

*Figure 14: Residual and Quantile Quantile Plots*

Residual Plots and Quantile Quantile (QQ) Plots are used to check the normality assumption used. The residual is the difference between the predicted value and the observed value of the dependent variable. Ideally, the residuals should be randomly distributed around the horizontal axis marking $y = 0$. Similarly, a straight line on the QQ plot denotes that the assumption is good enough to move ahead with.

## 3. Results (and discussions)

**Microbrewery**

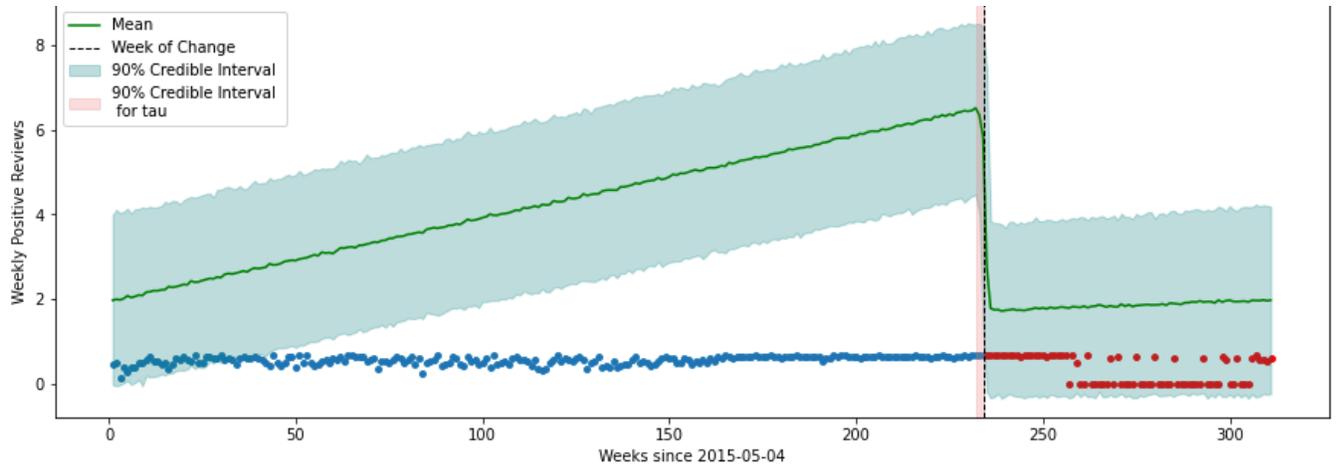

*Figure 15: Plot of the number of positive reviews against the week number for the Microbrewery category*

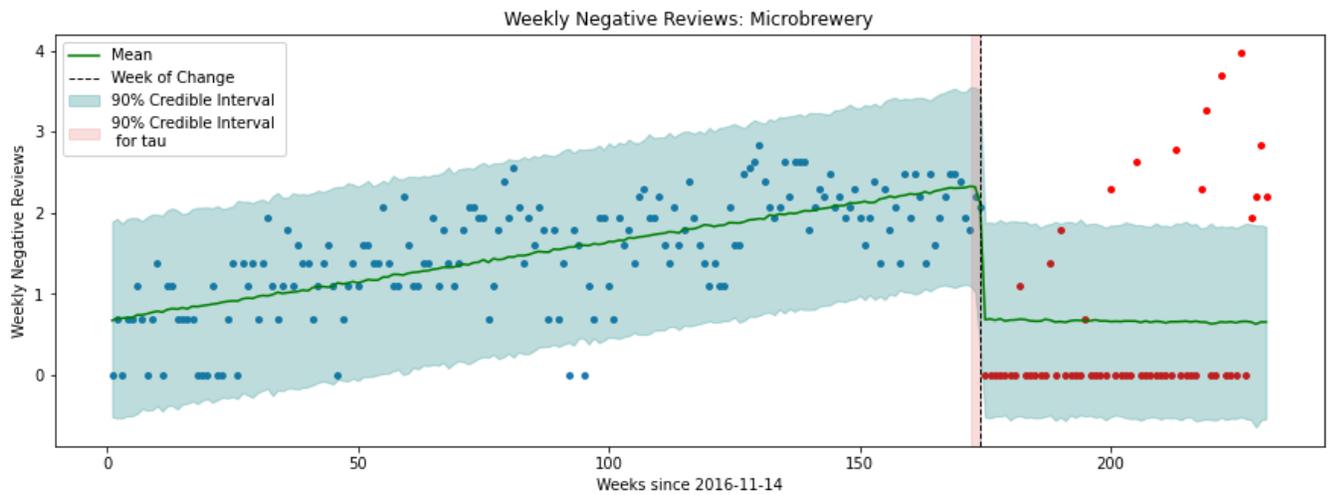

*Figure 16: Plot of the number of negative reviews against the week number for the Microbrewery category*

In Figure 15 and Figure 16, after the change point, a major dip is observed in the number of reviews with the 90% credible interval dropping down significantly. This tells us that the number of positive reviews has reduced significantly after the change point that was observed on **2019-10-28**. Since the change point is just before the beginning of the COVID-19 phase, we deduce that, in the COVID phase, the number of positive reviews has decreased considerably when compared to the pre-covid phase.

Now, in Figure 16, after the change point, we see that around 10% of the values of the number of negative reviews that are outside the 90% credible interval are on the higher scale of the values when compared to the pre-covid phase. Also, the change point was observed on **2020-03-16** which belongs to the COVID-19 phase. This indicates that the number of negative reviews for around 10% of the points after the change point is considerably higher than that observed in the pre-covid phase.

This tells us that COVID-19 is a change point in itself which has been proved through our Bayesian change point estimation method. Practically, the category of microbrewery has been seen to be affected exactly at the change point that we have found out from our method. Microbrewery here consists of the pubs in the Bangalore region and we know that the COVID-19 lockdowns severely impacted the business of these breweries.

**Bar**

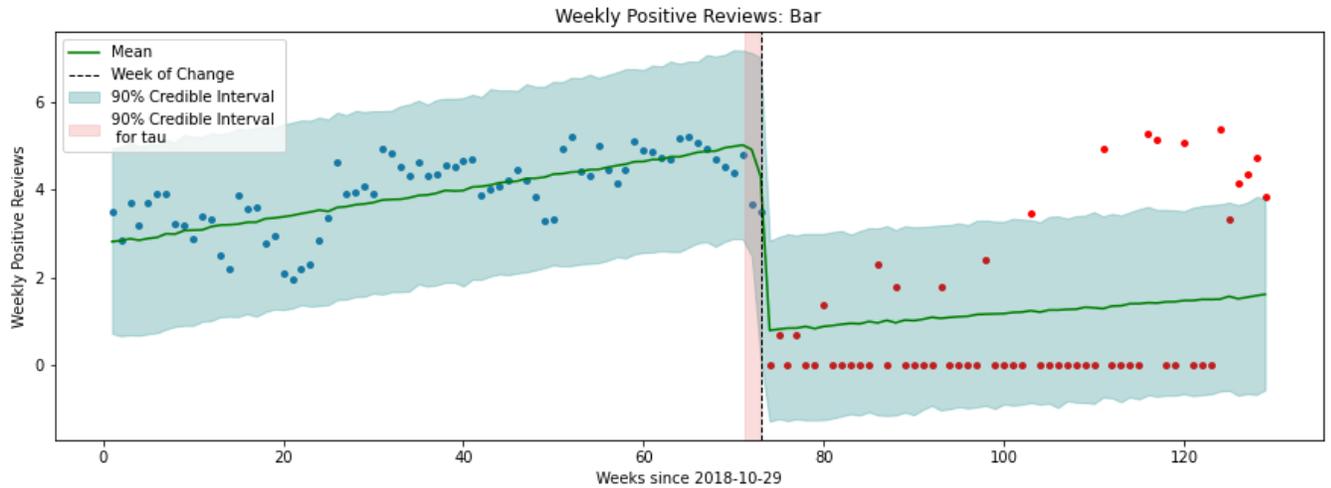

*Figure 17: Plot of the number of positive reviews against the week number for the Bar category*

In Figure 17, after the change point, a major dip is observed in the number of positive reviews with the 90% credible interval dropping down significantly. Since the change point is observed in the COVID-19 phase i.e. on **2020-03-23**, we deduce that, in the COVID phase, the number of positive reviews has decreased considerably when compared to the pre-covid phase. This is very intuitive since bars were shut down completely in the lockdown period. In Figure 17, we also observe that after the change point, around 10% of the points (that are outside the 90% credible interval of the number of positive reviews) are on a comparable scale to the points in the before-COVID phase. Further research is required to find reasons for the same.

**Microbrewery and Casual Dining**

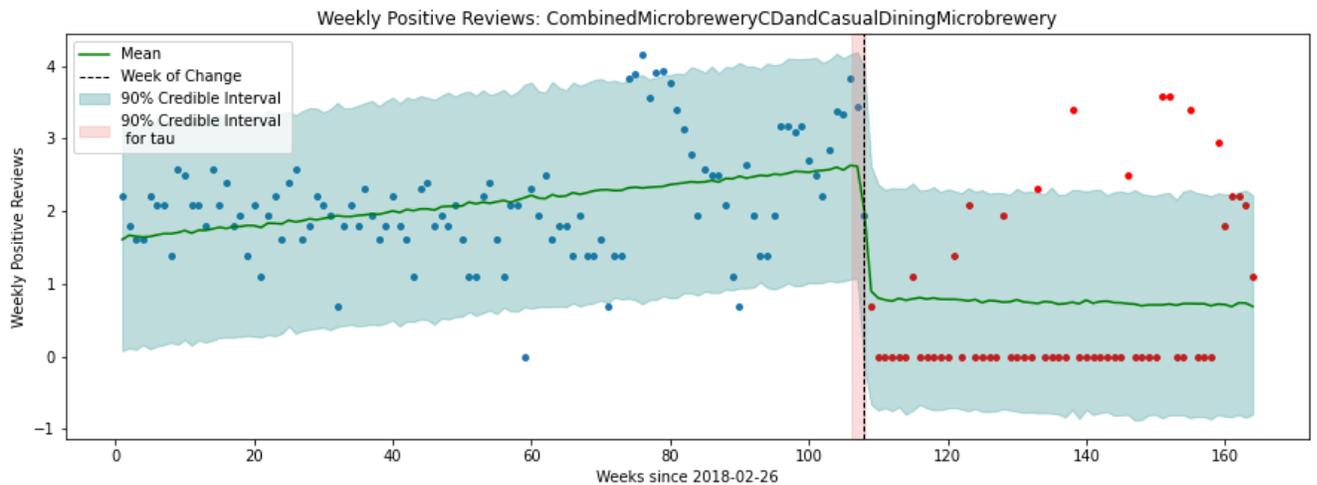

*Figure 18: Plot of the number of positive reviews against week number for the Microbrewery and Casual Dining category*

In Figure 18, after the change point, a major dip is observed in the number of positive reviews with the 90% credible interval dropping down significantly. Since the change point is observed in the COVID-19 phase, i.e. on **2020-03-23** we deduce that, in the COVID phase, the number of positive reviews has decreased considerably when compared to the pre-covid phase. In Figure 18, we also observe that after the change point, around 10% of the points (that are outside the 90% credible interval) are on a higher scale than the points in the before covid phase. Further research is required to find reasons for the same.

**Bar and Casual Dining**

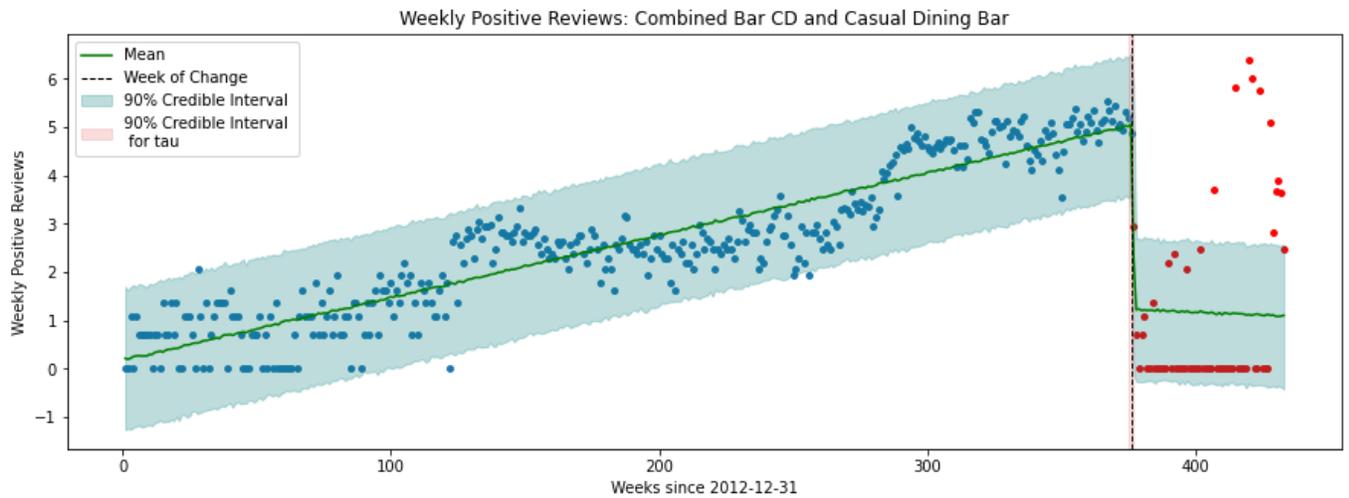

*Figure 19: Plot of the number of positive reviews against the week number for the Bar and Casual Dining category*

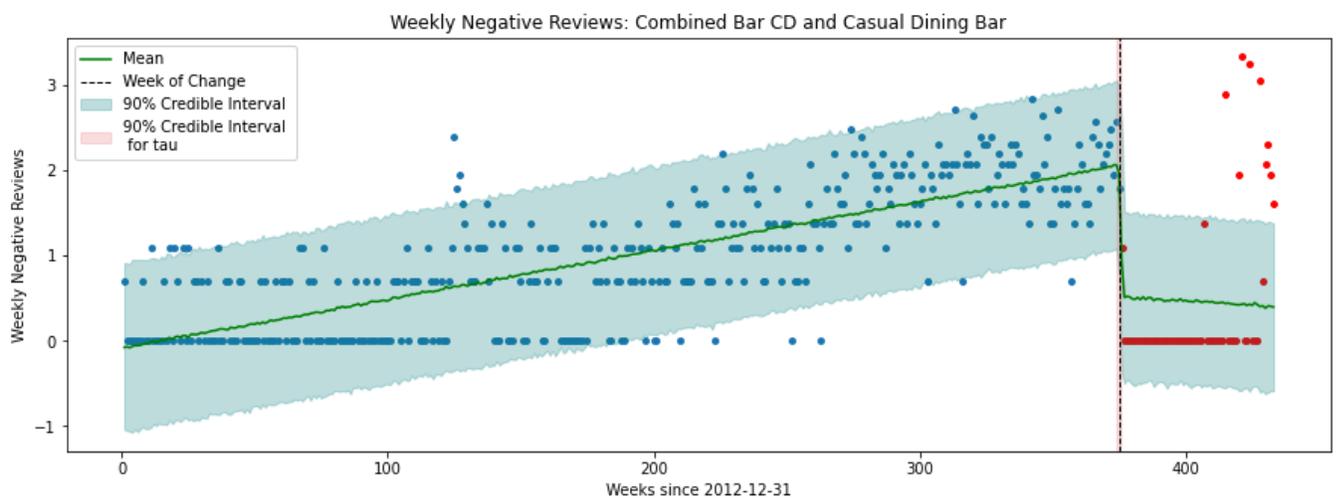

*Figure 20: Plot of the number of negative reviews against week number for the Bar and Casual Dining category*

In Figure 19 and Figure 20, after the change point, a major dip is observed in the number of reviews with the 90% credible interval dropping down significantly. Since the change point is observed in the COVID-19 phase, i.e. on **2020-03-09** we deduce that, in the COVID phase, the number of positive reviews has decreased considerably when compared to the pre-covid phase. In Figure 19, we also observe that after the change point, around 10% of the points (that are outside the 90% credible interval) are on a higher scale than the points in the before covid phase. Further research is required to find reasons for the same.

Now, in Figure 20, after the change point, we see that around 10% of the values of the number of negative reviews that are outside the 90% credible interval are on the higher scale of the values when compared to the pre-covid phase. Also, the change point was observed on **2020-03-16** which belongs to the COVID-19 phase. This indicates that the number of negative reviews for around 10% of the points after the change point is considerably higher than that observed in the pre-covid phase.

**Quick Bites Cafe**

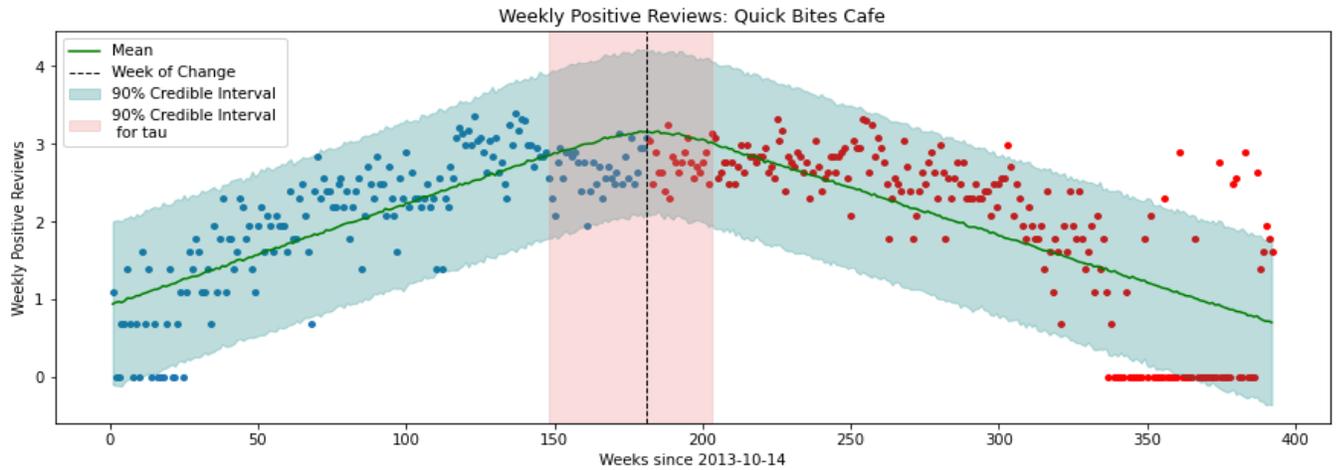

*Figure 21: Plot of the number of positive reviews against week number for the Quick Bites Cafe category*

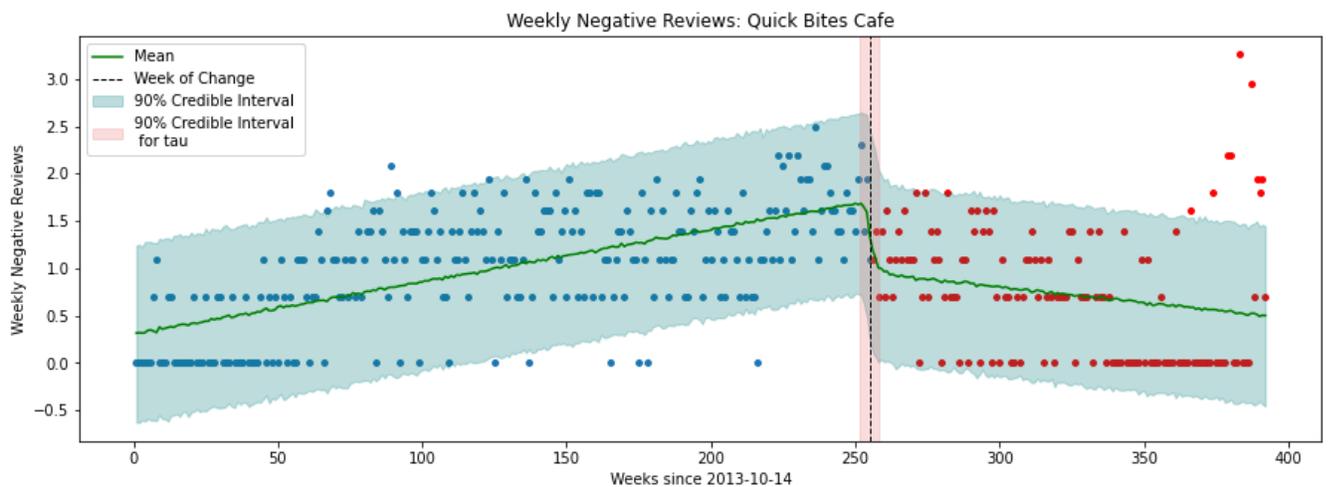

*Figure 22: Plot of the number of negative reviews against week number for the Quick Bites Cafe category*

In Figure 21 and Figure 22, after the change point, we observe a major dip in the 90% credible interval for the number of reviews. This tells us that the number of reviews, both positive and negative, has reduced considerably after the change point. The change point for the positive and negative reviews is **2017-04-03** and **2018-09-03** respectively outside the COVID-19 phase. Hence, the abrupt change in the number of reviews is not due to the pandemic, COVID-19. The results are very intuitive because essential services were not closed during the lockdown and hence the category is not expected to be impacted by COVID-19.

Table 2: Results for all the categories have been summarised in the table. Highlighted in green indicates a change point observed in the COVID-19 phase whereas highlighted in red indicates a change point not observed in the COVID-19 phase.

| Category Of Restaurant | Changepoint observed (yyyy-mm-dd) | Type of review | Comments |
| --- | --- | --- | --- |
| Microbrewery | **2019-10-28** | Positive | |
| Microbrewery | **2020-03-16** | Negative | |
| Bar | **2020-03-23** | Positive | |
| Casual Dining and Microbrewery | **2020-03-23** | Positive | |
| Casual Dining and Bar | **2020-03-16** | Positive | |
| Casual Dining and Bar | **2020-03-09** | Negative | The change point is observed in the covid phase. Also, the number of reviews increased till the change point and then started decreasing. |
| Casual Dining | **2020-03-16** | Positive | |
| Casual Dining | **2020-03-23** | Negative | |
| Quick Bites Cafe | 2017-04-03 | Positive | The change point observed is not in the covid phase. Also, the number of reviews increased till the change point and then started |
| Quick Bites Cafe | 2018-09-03 | Negative | |

| Quick Bites | 2016-08-22 | Positive | decreasing. |
| Casual Dining Lounge | Not Observed | Positive | Though the MCMC sampler converged, the r_hat value was found to be greater than the threshold value i.e. 1.1. |
| Casual Dining Lounge | Not Observed | Negative | |

**4. Conclusion & Future Work**

The COVID-19 pandemic has brought unprecedented challenges to the global restaurant industry, significantly impacting consumer behavior and business operations. This study, through its innovative application of Bayesian regression and predictive modeling, particularly the use of Hamiltonian Monte Carlo, has identified crucial change points in consumer preferences and behaviors towards different restaurant types. Our findings reveal that **nearly 4 out of 6 categories of restaurants had a pandemic impact with the change-point being detected strongly**. These insights are particularly valuable for restaurant owners and policymakers in strategizing for resilience and recovery in the post-pandemic era. Future research could build upon this study by how the posterior distribution of parameters can be used to determine the correctness of the change point observed. A threshold value for the overlap between posterior parameter plots will help us eliminate false positives.

**5. Acknowledgments**

**6. Declaration of interest statement**

I hereby declare that the disclosed information is correct and that no other situation of real, potential or apparent conflict of interest is known to me. Complete research is performed on the open-source data without any copyright issues.

[29] The Role of Technology in the Restaurant Industry Post-COVID-19. Available at: https://www.emerald.com/insight/content/doi/10.1108/JHTT-06-2020-0132/full/html

[30] Zupo, R., et al., 2020. Exploring the Impact of COVID-19 on the Sustainability of the Restaurant Industry. Available at: https://www.mdpi.com/1660-4601/17/19/7073

[31] Agarwal, R. (2019, June 4). MCMC Intuition for Everyone. *TowardsDataScience*. https://towardsdatascience.com/mcmc-intuition-for-everyone-5ae79fff22b1
Alagh, V., (2020, July 6). How Pandemic Is Reshaping Consumer Behaviour Post-Covid-19?. *Inc 42*. https://inc42.com/resources/how-pandemic-is-reshaping-consumer-behaviour-post-covid-19/

[31] Ali, A. (2021, October 26). 25% eateries closed in FY21 due to Covid: Study. *Times of India*. https://timesofindia.indiatimes.com/india/25-eateries-closed-in-fy21-due-to-covid-study/articleshow/87268787.cms, para.1)

[32] Aminikhanghahi, S., J. Diane, C.J. (2016). A Survey of Methods for Time Series Change Point Detection. PubMed Central. 51(2): 339–367.

[33] Becker, C. (2009), "How a natural disaster changed the face of the restaurant industry in southern Mississippi: A Katrina story", Journal of Foodservice Business Research, Vol. 12 No. 3, pp. 266-274.

[34] Chhetri, S., Dambhare, A., & Seth, P. (2021). Analysing the impact of COVID-19 on consumer behaviour towards the hotel industry in India. Research Gate. https://www.researchgate.net/publication/353018698_ANALYSING_THE_IMPACT_OF_COVID-19_ON_CONSUMER_BEHAVIOUR_TOWARDS_HOTEL_INDUSTRY_IN_INDIA

[35] Clyde, M., Çetinkaya-Rundel, M., Rundel, C., Banks, D., Chai, C., Huang, L.. (2019) An Introduction to Bayesian Thinking- A Companion to Statistics with R Course. Available online at: https://statswithr.github.io/book/bayesian-inference.html#credible-intervals-and-predictive-inference

[36] DAC team. (2020, April 13). How the pandemic is changing consumer behavior. *Covid-19 Digital Communication Series*. https://www.dacgroup.com/blog/how-the-pandemic-is-changing-consumer-behavior/

[37] Elster, C., Klauenberg, K., Walzel, M., W¨ubbeler, G., Harris, P., Cox, M., Matthews, C., Smith, I., Wright, L., Allard, A., Fischer, N., Cowen, S., Ellison, S., Wilson, P., Pennecchi, F., Kok, G., Veen, A.V.D., & Pendril, l.L. (2015). A Guide to Bayesian Inference for Regression Problems.

[38] Gundersen, G. (2019, August 13). Bayesian Online Changepoint Detection. *GregoryGunderson*. http://gregorygundersen.com/blog/2019/08/13/bocd/
Jia, S. (Sixue). (2021). Analyzing Restaurant Customers' Evolution of Dining Patterns and Satisfaction during COVID-19 for Sustainable Business Insights. *Sustainability*, *13*(9), 4981. MDPI AG. Retrieved from http://dx.doi.org/10.3390/su13094981